\documentclass[refree]{raa}
\usepackage{graphicx,times}
\usepackage{natbib}
\usepackage{amssymb,amsmath}
\bibpunct{(}{)}{;}{a}{}{,}
\usepackage[colorlinks=true,linkcolor=green,citecolor = blue]{hyperref}%
\makeatletter
\newcommand*{\rom}[1]{\expandafter\@slowromancap\romannumeral #1@}
\makeatother

\def\fm{\hbox{$.\!\!^m$}}
\def\fs{\hbox{$.\!\!^s$}}
\def\degr{\hbox{$^\circ$}}

\begin{document}%

\title{Orbital and physical parameters of the close binary system: GJ 9830 (HIP\,116259)}

\volnopage{ {\bf 2019} Vol.\ {\bf } No. {\bf0}, 000--000}
   \setcounter{page}{1}

   \author{ Suhail G. Masda\inst{1,2}, Mashhoor A. Al-Wardat\inst{3} and J. M. Pathan \inst{4}}

\institute{  Physics Department, Dr. Babasaheb Ambedkar Marathwada University, Aurangabad-431001, Maharashtra, India; \textit{suhail.masda@gmail.com} \\
		 \and Physics Department, Hadhramout University, PO Box:50511, Mukalla, Yemen \\
	     \and Physics Department, Al al-Bayt University, PO Box: 130040, Mafraq, 25113 Jordan\\
	     \and Physics Department, Maulana Azad College, Aurangabad-431001, Maharashtra, India\\
\vs \no
   {\small Received 2019 Jan 24; accepted 2019 Feb 22}
}

\abstract{
We present the complete set of physical and geometrical parameters of the visual close binary system GJ\,9830 for the first time by using Al-Wardat's complex method. This method combines  magnitude difference from speckle interferometry, synthetic spectral energy distributions of the binary components which are constructed depending on grids of Kurucz blanketed models (Atlas9), along with the orbital solution by using Tokovinin's dynamical method to estimate the parameters of the individual components.
The analysis of the system by using synthetic photometry resulted in the following set of parameters: $T_{\rm eff.}=6220\pm 100$ \,K, $\rm log~g=4.30\pm 0.12$, $R=1.10\pm0.08\,R_\odot$ for the primary component  and  $T_{\rm eff.}=4870\pm 100$\,K, $\rm log~g=4.60\pm 0.11$, $R=0.709\pm0.07\,R_\odot$ for  the secondary component. The recently published dynamical parallax from \textit{Gaia} space mission was used to calculate the total mass of the binary system as $1.75\pm0.06\, \mathcal{M}_\odot$ which coincides with those estimated by using Al-Wardat's method as $ \mathcal{M}^{A}=1.18\pm0.10\, \mathcal{M}_\odot$, $ \mathcal{M}^{B}=0.75\pm0.08\, \mathcal{M}_\odot$. 
 The  analysis of the system reveals that both components belong to main sequence stars with an age around  $1.4\pm0.50$\,Gyr. The evolutionary tracks and isochrones of the system's components are discussed, and the fragmentation process is suggested as the most likely process for the formation of the system. 
}

\authorrunning{Masda et al. }            
   \titlerunning{Parameters of GJ\,9830 (HIP\,116259)}  
\maketitle%

\section{INTRODUCTION }
 The physical and geometrical parameters especially in the close binary systems  play a definitive role in understanding more problems in formation and evolution for those binaries. One of those problems is the stellar  mass which gives insight into evolution of the binary systems. Precise parallax of the binary system especially  from \textit{Gaia}
 astrometric mission \cite{2018yCat.1345....0G} plays a vital role in enhancing the value of absolute
 magnitudes and binary orbits with reliable stellar masses.

Speckle interferometry \citep{2002AA...385...87B,2010AJ....139..743T,2011AJ....142..176M} and adaptive
optics  \citep{2005AJ....130.2262R,2011AJ....142..175R} are modern high-resolution techniques of ground based observations and instrumental in resolving the  close visual binary systems. Speckle interferometry is a significant technique for the study of the visual and spectroscopic binary stars
 which is used to measure the position angles ($\theta$), separation anagles ($\rho$) and magnitude differences ($\triangle m$) for the subcomponents of the binary and multiple system \citep{2002AA...385...87B,2010AJ....139..743T}. Most of analytic methods have been used results of this technique to estimate the orbital and physical parameters such as Kowalsky's method ~\citep{1930MNRAS..90..534S}, Fourier transform method \citep{1979ApJ...234..275M}, Tokovinin's dynamical method \citep{1992ASPC...32..573T}, Koval'skij method \citep{2004A&A...415..259O}, Docobo's analytic method \citep{1985CeMec..36..143D,2012ocpd.conf..119D,2018MNRAS.476.2792D} and Al-Wardat's complex method \citep{2007AN....328...63A,2014AstBu..69..198A,2017AstBu..72...24A}, respectively.

 In our analysis, we make use of Tokovinin's dynamical method to estimate orbital parameters \citep{1992ASPC...32..573T} by using a modern version of Tokovinin's ORBITX program. The method requires the knowledge of $\theta$, $\rho$ and the epoch of the orbit \citep{2017AJ....154..110T,2018yCat..51540110T}.

 On the other hand, the physical parameters are of fundamental value in terms of testing the formation and evolution models of the binary system beside the orbital solution. As a result, we follow Al-Wardat's complex method \citep{2007AN....328...63A} which combines the results of spectrophotometry with the results of speckle interferometry to obtain the physical parameters \citep{2012PASA...29..523A,2014AstBu..69..198A,2016RAA....16..166A,2016RAA....16..112M,2018JApA...39...58M}. This method makes use of the entire spectral energy distributions (SEDs)
 of the binary systems which construct by using Atlas9 atmospheric modelling \citep{1994KurCD..19.....K}.

 In addition, synthetic photometry is used to estimate the physical parameters
 more accurately through the colour indices without needing the observed spectra of the binary systems \citep{1996BaltA...5..459S,1999A&A...346..564C,2012PASP..124..140B,2013AJ....146...68L}. It is a quantitatively analysis for the synthetic SED of a binary system which is about modifying stellar parameters
 such that the synthetic magnitudes fit the observed ones \citep{2014AstBu..69..198A,2014AstBu..69...58A,2016RAA....16..112M,2017AstBu..72...24A}. The method, which should be followed to evaluate the stellar parameters of the binary system by using synthetic photometry throughout the analysis of the system was described in evident detials in previous paper \citep{2018JApA...39...58M}.
 
 Two methods have been successfuly applied  to estimate the physical and geometrical parameters to several solar-type stars and sub-giant binary stars whether the observed spectra were avaliable or not such as HD\,25811, HD\,375, Gliese\,762.1, FIN\,350, COU1511, HIP105947 and two systems HIP\,14075 and HIP\,14230 ~\citep{2014AstBu..69...58A,2014PASA...31....5A,2016RAA....16..166A,2017AstBu..72...24A,2018RAA....18...72G,2018JApA...39...58M}.

 The system GJ9830 (HIP116259) is a well-known close binary system in the solar neighborhood. This system located at the \textit{Gaia} parallax of 29.178$\pm$0.186 mas \citep{2018yCat.1345....0G} which is implied to a precise kinematic distance of  34.27$\pm$0.0002 pc. \cite{2007AstBu..62..339B} estimated close binary system orbital solution for GJ9830 by using the \cite{1977ApJ...214L.133M} method. Their estimated total mass was $1.56\pm0.18\rm\, \mathcal{M}_\odot$  under the old \textit{Hipparcos} parallax of $30.24$\,mas~\citep{1997yCat.1239....0E}.
 The last observed relative position measurement applied by \cite{2007AstBu..62..339B}  was at epoch 2006.946. As a result, eightteen new interferometric measurements from epoch 2002 to epoch 2011 are included in our orbit (Table~\ref{tab}). Due to the changes in residuals especially in $\rho$, it was necessary to obtain a new orbit.

 Our main aim is to estimate the orbital and physical parameters of the close binary system GJ\,9830 by using Tokovinin's dynamical and Al-Wardat's complex method, respectively. Moreover, we employed the new parallax of the system from \textit{Gaia} space mission.

\section{Observational data}
Our study depends on observational photometric data which are taken from reliable different sources such as Hipparcos \cite{1997yCat.1239....0E}, Str\"{o}mgren \citep{1998A&AS..129..431H} and TYCHO
catalogues \citep{2000A&A...355L..27H}. These data are used as reference and comparison with synthetic photometric results to get the best stellar parameters of the system. In addition to that, we have obtained new data from interferometric measurements of the system for the sake of reconstructing the system's orbit (Table~\ref{tab}).

Table~\ref{tabl0} contains fundamental  and the observed photometric data for GJ\,9830 from SIMBAD database, NASA/IPAC, Str\"{o}mgren, the Hipparcos and TYCHO
catalogues.

\begin{table}[ht]
	\centering
	\caption{Fundamental and observed photometric data for the system GJ 9830.} \label{tabl0}
	\begin{tabular}{ccc}\hline\hline
		Property & GJ9830 &  Ref.  \\
		\hline			\noalign{\smallskip}
		$\alpha_{2000}$ $^{a}$  & $23^h 33^m 24\fs06$ &  simbad\\
		$\delta_{2000}$ $^{b}$  & $+42\degr50' 47.''86$	&  -\\
		Sp. Typ.  &  G0  &  -
		\\
		E(B-V)& $ 0.099\pm0.01 $ &  $c$
		\\		
		$A_v$   &  $0\fm30$  & $c$
		\\
		$\pi_{Hip}$ (mas) &    $30.24\pm1.12$      &  $d$
		\\
		$\pi_{Hip}$  (mas) &    $25.04\pm0.74$   & $e$\\
		$\pi_{Gaia}$  (mas) &    $29.178\pm0.186$   & $f$\\
		$V_J$  & $7\fm14$  &  $d$
		\\
		$B_J$   &  $7\fm72$  &   $g$
		\\
		$(V-I)_J$& $0\fm79\pm0.01$ &  $d$
		\\				
		$(B-V)_J$& $0\fm585\pm0.008$ & -
		\\
		$(b-y)_S$& $0\fm40\pm0.002$ &  $h$
		\\
		$(v-b)_S$& $0\fm58\pm0.002$ &  -\\
		$(u-v)_S$& $0\fm86\pm0.007$ &  -\\
		
		$B_T$  &   $7\fm86\pm0.007$  &  $g$
		\\			
		$V_T$ &   $7\fm23\pm0.006$    & -
		\\
		\hline\hline
	\end{tabular}
	
	\textbf{Notes.} $^{a}$ Right Ascension and $^{b}$ Declination.\\
		$^{c}$ \cite{2011ApJ...737..103S}, $^d$ Old \textit{Hipparcos} \citep{1997yCat.1239....0E}, $^e$ New \textit{Hipparcos} \cite{2007A&A...474..653V}, $^f$\citep{2018yCat.1345....0G}, $^g$ \citep{2000A&A...355L..27H} and $^h$ \citep{1998A&AS..129..431H}.
\end{table}

\section{Method and Analysis}

\subsection{Orbital parameters}\label{22}
\begin{figure}
	\centering
	\includegraphics[angle=0,width=12cm]{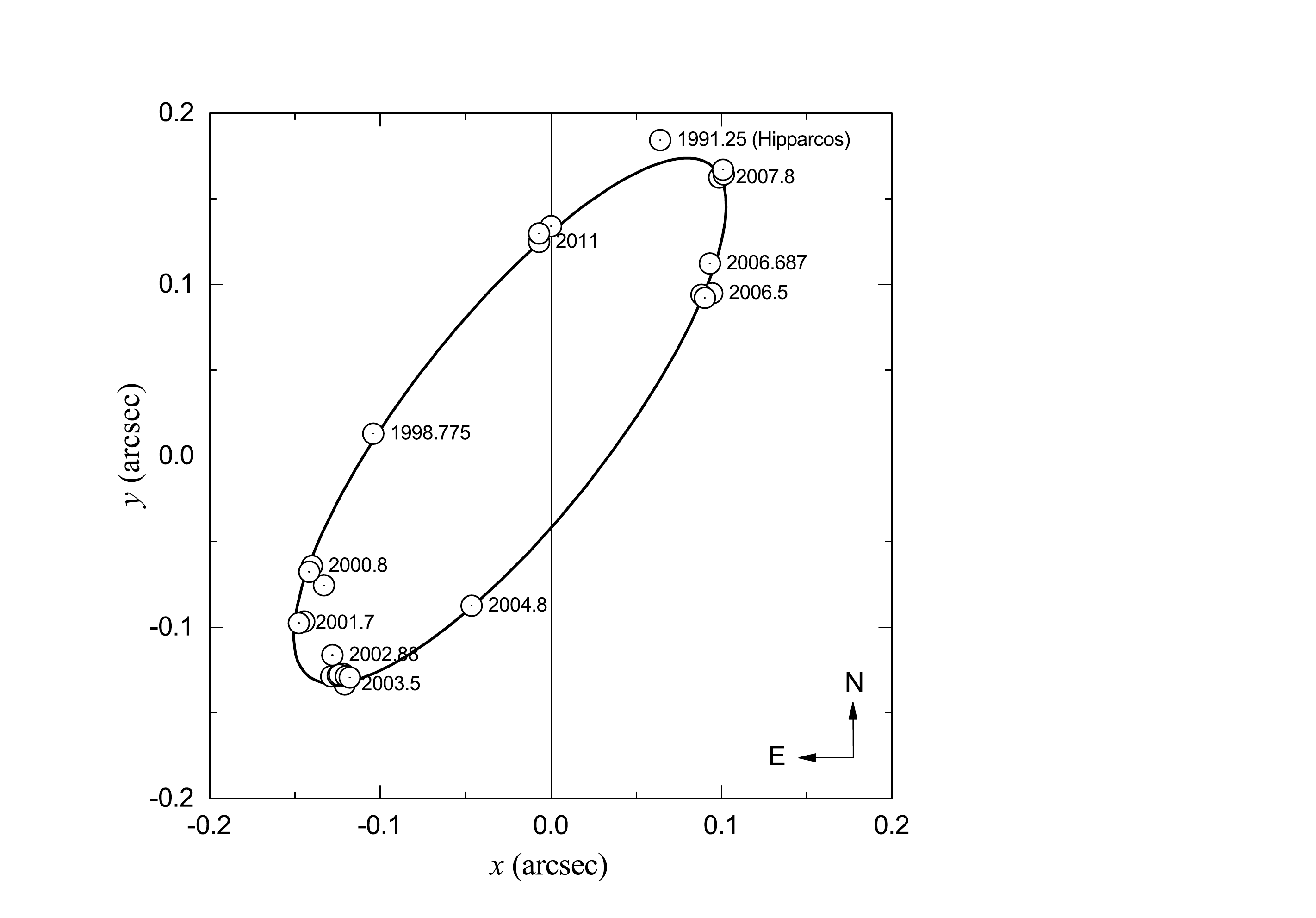}
	\caption{The modified apparent orbit of the GJ 9830
		binary system calculated by using the interferometric measurements
		from the INT4 (with 18 new measurements). The origin point represents the position of the primary component.
	} \label{figor}
\end{figure}

Understanding of the relative motion of the secondary star around the primary  star of a binary system is essentially a matter of determining the orbital parameters. As a result, we follow Tokovinin's dynamical method to do so and use the ORBITX code of \cite{1992ASPC...32..573T} to get the best orbit. The angular separations ($\rho$) and position angles ($\theta$) are obtained in Table~\ref{tab} from INT4. The program
performs a least-squares adjustment to all available relative position observations, with weights inversely proportional to the square of their standard errors. The orbit solution involves: the orbital period, P; the eccentricity, e; the semi-major axis, a; the inclination, $ \ i$; the argument of periastron, $ \omega$; the position angle of nodes, $\Omega$; and the time of primary minimum, $\ T_0$. Hence, the modified orbit is shown in Figure \ref{figor} and the orbital parameters are listed in Table~\ref{tabb}.

\begin{table*}[ht]
	\centering
	\caption{Relative position measurements, residuals $\triangle\theta $ and $\triangle\rho $ (our work)  of the system, which are used to build the orbit of the system.} \label{tab}
	\begin{tabular}{cccccc}\hline\hline
		Data  & $\theta$ & $\rho$ &  $\triangle\theta $  & $\triangle\rho $ & Ref.
		\\
		Epoch  & (\degr)& ($''$) & (\degr) & ($''$) & \\
		\hline \noalign{\smallskip}
		1991.25 & $341.0$  & $0.195$&   15.3  & 0.013 & \cite{1997ESASP1200.....E} \\
		1998.7764  & $83.0$ & $0.105$ &  -2.2   &  -0.000 & \cite{2002AA...385...87B}\\
		2000.6171 & $119.6$&  $0.153$  &  6.0 & -0.001 & \cite{2002AJ....123.3442H} \\
		
		2000.7590 & $114.7$  &  $0.154$ & -0.3  &  -0.004  &\cite{2002AJ....123.3442H}\\
		2000.8646 & $115.6$  &$0.157$  &  -0.5 & -0.003  & \cite{2006BSAO...59...20B} \\
		2000.8727 & $115.6$  &$0.157$  &  -0.5 &  -0.004 & \cite{2006BSAO...59...20B} \\
		2001.7607 & $123.8$  &$0.174$  & 0.2  &  -0.007  & \cite{2006BSAO...59...20B} \\	
		2001.7607 & $123.5$  &$0.177$  &  -0.1 &  -0.004 & \cite{2006BSAO...59...20B} \\
		
		2002.8820$^*$ & 132.28 & 0.173 &  0.9& -0.019 & \cite{2009ApJS..181...62M}\\
		
		2003.5304$^*$ & 137.90 & 0.180 & 2.1  & -0.005 &\cite{2008AJ....136..312H}  \\
		2003.5304$^*$  & 135.0 & 0.182 &  -0.8 &  -0.003 &\cite{2008AJ....136..312H}\\
		2003.5386$^*$ & 136.4 & 0.176 & 0.6   & -0.009 &\cite{2008AJ....136..312H}\\

		2003.5386$^*$  & 135.8 & 0.179 & -0.0 & -0.006 &\cite{2008AJ....136..312H}\\
		2003.5386$^*$ & 135.9 & 0.178 & 0.1 & -0.007&\cite{2008AJ....136..312H}\\
		2003.6343$^*$  & 137.0 & 0.176 & 0.5 & -0.007 &\cite{2008AJ....136..312H}\\
		2003.6343$^*$  & 137.7 & 0.175 & 1.2 & -0.008 &\cite{2008AJ....136..312H}\\

		2004.8240  & 152.1 & 0.099 & 3.1 & -0.017 & \cite{2007AstBu..62..339B}\\
		
		2006.5174$^*$  & 315.3 & 0.134 & 1.3 & 0.018 & \cite{2008AJ....136..312H}\\
		2006.5202$^*$ & 316.9 & 0.129 & 2.9   & 0.012 & \cite{2008AJ....136..312H} \\
		2006.5256$^*$ & 315.7 & 0.129 & 1.6  & 0.012 & \cite{2008AJ....136..312H} \\
		2006.6870$^*$ & 320.4 & 0.146 & 3.7  & 0.014 &\cite{2013AstBu..68...53B}\\
		
		
		2007.817$^*$ & 328.9 & 0.190 & 1.8 &  0.003 &\cite{2010AJ....139..205H}\\
		2007.8201$^*$  & 328.5 & 0.193 & 1.4  &  0.006 & \cite{2010AJ....139..205H}\\
		2007.8253$^*$ & 329.0 & 0.195 & 1.8  &  0.008 & \cite{2010AJ....139..205H}\\
		2011.6837$^*$  & 0.00 & 0.1344 & 0.0   &  0.004 & \cite{2017AJ....153..212H} \\
		2011.9402$^*$  & 3.20 & 0.1245 & -0.7  & 0.001 &\cite{2017AJ....153..212H} \\
		2011.9402$^*$  & 3.00 & 0.1301 & -0.9  &  0.007 & \cite{2017AJ....153..212H} \\
		\hline\hline
	\end{tabular}\\
	$^*$ New data from interferometric measurements of the GJ 9830 binary system.
\end{table*}

We calculate the total mass and the the corresponding error of the binary system by using Kepler's third law and employing the estimated orbital parameters, semi-major axis in arcseconds, orbital period in years (see Table~\ref{tabb}) and the new parallax from~\cite{2018yCat.1345....0G} in arcseconds, as follows:
\begin{eqnarray}
	\label{eq31}
	\mathcal{M}_{T}=(\frac{a^3}{\pi^3P^2})\ \mathcal{M}_\odot
\end{eqnarray}
\begin{eqnarray}
	\label{eq32}
	\frac{\sigma_\mathcal{M} }{\mathcal{M}} =\sqrt{(3\frac{\sigma_\pi}{\pi})^2+(3\frac{\sigma_a}{a})^2+(2\frac{\sigma_p}{p})^2}
\end{eqnarray}
This equation gives the total mass and the corresponding error of the binary system  as $1.75\pm0.06\rm\, \mathcal{M}_\odot$. This result will subsequently be compared with the estimated theoretical individual masses of the binary system from  Al-Wardat's  complex method.

\subsection{Physical parameters}\label{10}
Estimating the physical parameters of the binary system GJ9830 by using Al-Wardat's  complex method for analysing VCBSs needs a determination of the magnitude difference between the components of the system properly. So, we estimate the visual magnitude difference of the system as $\triangle m=2.47\pm0.07$ mag between the two components as the average for all $\triangle m$ measurements given in Table~\ref{taba1} under the 541-562 nm V-band filters. Combining that value with the entire visual magnitude of the system obtains the apparent visual
magnitudes of individual components as: $m_v^A=7\fm25\pm0.08$ and  $ m_v^B=9\fm72\pm0.21$ for the primary and secondary components of the system, respectively

\begin{table}[ht]
	\begin{center}
		\caption{Speckle interferometric magnitude differences and
			Hipparcos $ \Delta H_{Hip}$ measurements of the system, along with filters used to obtain the observations from Fourth Catalog of Interferometric Measurements of Binary Stars (INT4).}
		\label{taba1}
		\begin{tabular}{c|cccc}
			\noalign{\smallskip}
			\hline\hline
			HIP	&$\triangle m $& {$\sigma_{\Delta m}$}& Filter ($\lambda/\Delta\lambda$)& Ref.  \\
			\hline
			GJ 9830	&	$2\fm61$ &   0.58  & $V_{Hip}:545nm/30$&  1   \\
			
			&	$2\fm48$ &   0.04  & $545nm/30$& 2    \\		
			&	$2\fm53$ &   0.07  & $545nm/30$& 3    \\						
			&	$2\fm40$ &  0.12  & $648nm/41$& 4   \\
			&	$2\fm21$ &   0.12  &$648nm/41$& 4      \\
			&	$2\fm16$ &  0.03 &$600nm/30$  & 5    \\
			&	$2\fm18$ &  0.10 &$600nm/30$  & 5    \\
			&	$2\fm60$ &   *  &$550nm/40$& 6      \\
			&	$2\fm47$ & * &$541nm/88$  &  6    \\
			&	$2\fm45$ &  0.03  & $545nm/30$ & 7    \\
			&	$2\fm28$ &  *  &$550nm/40$& 6      \\
			&	$2\fm34$ &  * &$550nm/40$  & 6    \\
			&	$2\fm52$ &  * &$550nm/40$  & 8    \\
			&	$2\fm64$ &   *  &$550nm/40 $& 8     \\
				&	$2\fm26$ & *  & $562nm/40 $ & 9   \\
			\hline\hline
		\end{tabular}
		\\
			$^*$ indicates that these measurements have no errors in INT4.
		$^1$\cite{1997yCat.1239....0E},
		$^2$\cite{2005A&A...431..587P},
		$^3$\cite{2002AA...385...87B},
		$^4$\cite{2004AJ....127.1727H},
		$^5$\cite{2006BSAO...59...20B},
		$^6$\cite{2008AJ....136..312H},
		$^7$\cite{2007AstBu..62..339B},
		$^8$\cite{2010AJ....139..205H},
		$^9$\cite{2017AJ....153..212H}.
	\end{center}
\end{table}

The results of the apparent visual magnitudes, combined with the parallax from \textit{Gaia} \cite{2018yCat.1345....0G} of $29.178\pm0.186 $ mas, lead to the absolute visual
magnitudes for components of the system as $ M_V^A=4\fm58\pm0.21$  and  $M_V^B=7\fm05\pm0.28$ for the primary and secondary components of the system, respectively by using the following equation \citep{1978GAM....15.....H} (see p.28):
\begin{eqnarray}
	\label{eq3}
	\ M_V=m_v+5-5\log(d)-A_v
\end{eqnarray}
Here, the interstellar extinction  can be neglected because the studied binary system is a nearby star.

Errors of the absolute visual magnitudes of the components of the system in Equation~\ref{eq3} are calculated by using the following relation:

\begin{eqnarray}
	\label{eq341}
	\sigma_{M^{*}_{V}} =\pm \sqrt{\sigma_{m^{*}_{v}}^2+(\frac{5 \log e}{\pi_{Hip}})^2\sigma_{\pi_{Hip}}^2}~~~~~ *=A,B.
\end{eqnarray}
Here, the $ \sigma_{m^{*}_{v}} $ are the errors of the apparent visual magnitudes.

Based on the above estimated absolute magnitudes ($M_{V}$) of the individual components of the system  and their relations with effective temperatures ($T_{\rm eff.}$) in addition to Tables \citep{1992adps.book.....L,2005oasp.book.....G} and the below equations \ref{eq8} and \ref{eq5}, we obtain input preliminary parameters of the system as: $T_{\rm eff.}=5878K$, log g = 4.36, $R=1.10R_\odot$ for the primary component and $T_{\rm eff.}=4798K$, log g = 4.54, $R=0.74R_\odot$ for the secondary component.
\begin{eqnarray}
	\label{eq8}
	\log(R/R_\odot)= 0.5 \log(L/L_\odot)-2\log(T_{\rm eff.}/T_\odot)\\
	\label{eq5}
	\log g = \log(M/M_\odot)- 2\log(R/R_\odot) + 4.43,
\end{eqnarray}
Here
$T_\odot$ was taken as $5777\rm{K}$.

In order to test the results of the synthetic photometry of the system for the sake of obtaining the best stellar parameters, we need to construct the synthetic SED of the system based on the input parameters  and on grids of blanketed models (ATLAS9) \citep{1994KurCD..19.....K}. Hence, the entire synthetic SED at Earth of the binary system, which is connected to the individual synthetic SEDs of the binary system, is computed by using the following equation:
\begin{eqnarray}
	\label{eq7}
	F_\lambda  = (R_{A} /d)^2(H_\lambda ^A + H_\lambda ^B \cdot(R_{B}/R_{A})^2) ,
\end{eqnarray}
\noindent
where $ R_{A}$ and $ R_{B}$ are the radii of the primary and secondary components of the system in solar units, $H_\lambda ^A $ and  $H_\lambda ^B$ are the fluxes at the surface of the star and $F_\lambda$ is the flux for the entire SED of the system above  the Earth's atmosphere which is located at a revised distance d (pc) from the system.

As a result of the lack of knowledge of the observed spectrum of the binary system, we will depend on the results of the synthetic SED by using synthetic photometry for the sake of the reliability for accurate physical parameters. This technique is essentially dependent on the results of Al-Wardat's complex method. Our aim is to obtain the best agreement between the observed colour indices and magnitudes of the entire system with the entire synthetic SED of the system and consequently obtain the best physical parameters of the binary system.

\subsubsection{Synthetic photometry} \label{12}
The stellar parameters are mainly dependent on the best fit between the observed coulors indices and magnitudes of the entire system with the entire synthetic SED of the system whether the observed spectrum was available or not. Therefore, the entire and individual synthetic magnitudes and coulors indices of the binary system are calculated by integrating the model fluxes over each bandpass of the system calibrated to the reference star (Vega) by using the following equation \citep{2007ASPC..364..227M,2012PASA...29..523A}:

\begin{equation}\label{15}
	m_p[F_{\lambda,s}(\lambda)] = -2.5 \log \frac{\int P_{p}(\lambda)F_{\lambda,s}(\lambda)\lambda{\rm d}\lambda}{\int P_{p}(\lambda)F_{\lambda,r}(\lambda)\lambda{\rm d}\lambda}+ {\rm ZP}_p\
\end{equation}
where $m_p$ is the synthetic magnitude of the passband $p$, $P_p(\lambda)$ is the dimensionless sensitivity function of the passband $p$, $F_{\lambda,s}(\lambda)$ is the synthetic SED of the object and $F_{\lambda,r}(\lambda)$ is the SED of Vega.  Zero points (ZP$_p$) from~\cite{2007ASPC..364..227M} (and references there in) were adopted.

In order to obtain accurate physical parameters, it is necessary to have an accurate knowledge of the following criteria:
\begin{enumerate}
	\item The colour indices and magnitudes of the synthetic photometry should be computed of the studied binary system by using equation \ref{15}.
	\item The colour indices and magnitudes, \textit{$B-V$, $b-y$, $V_{J}$}, etc. of the entire SED should be completely consistent with observed ones of the binary system.
	\item The magnitude difference between the components  ($\triangle m=V^{b}_{J}-V^{a}_{J}$) of
	the synthetic photometry should be consistent with observed one.

\end{enumerate}

The color  indices of the binary system are the strong indication for the sake of reaching the best stellar parameters. So, under the prceeding criteria, the final results of the calculated magnitudes and color  indices within three different photometrical systems-Johnson: $U$, $B$, $ V$, $R$, $U-B$, $B-V$, $V-R$; Str\"{o}mgren: $u$, $v$, $b$,
$y$, $u-v$, $v-b$, $b-y$ and Tycho: $B_{T}$, $ V_{T}$, $B_{T}-V_{T}$)- of the entire
synthetic system and individual components of the system GJ 9830, are shown in Table~\ref{tab42}.

\begin{figure}[h]
	\centering
	\includegraphics[angle=0,width=8.3cm]{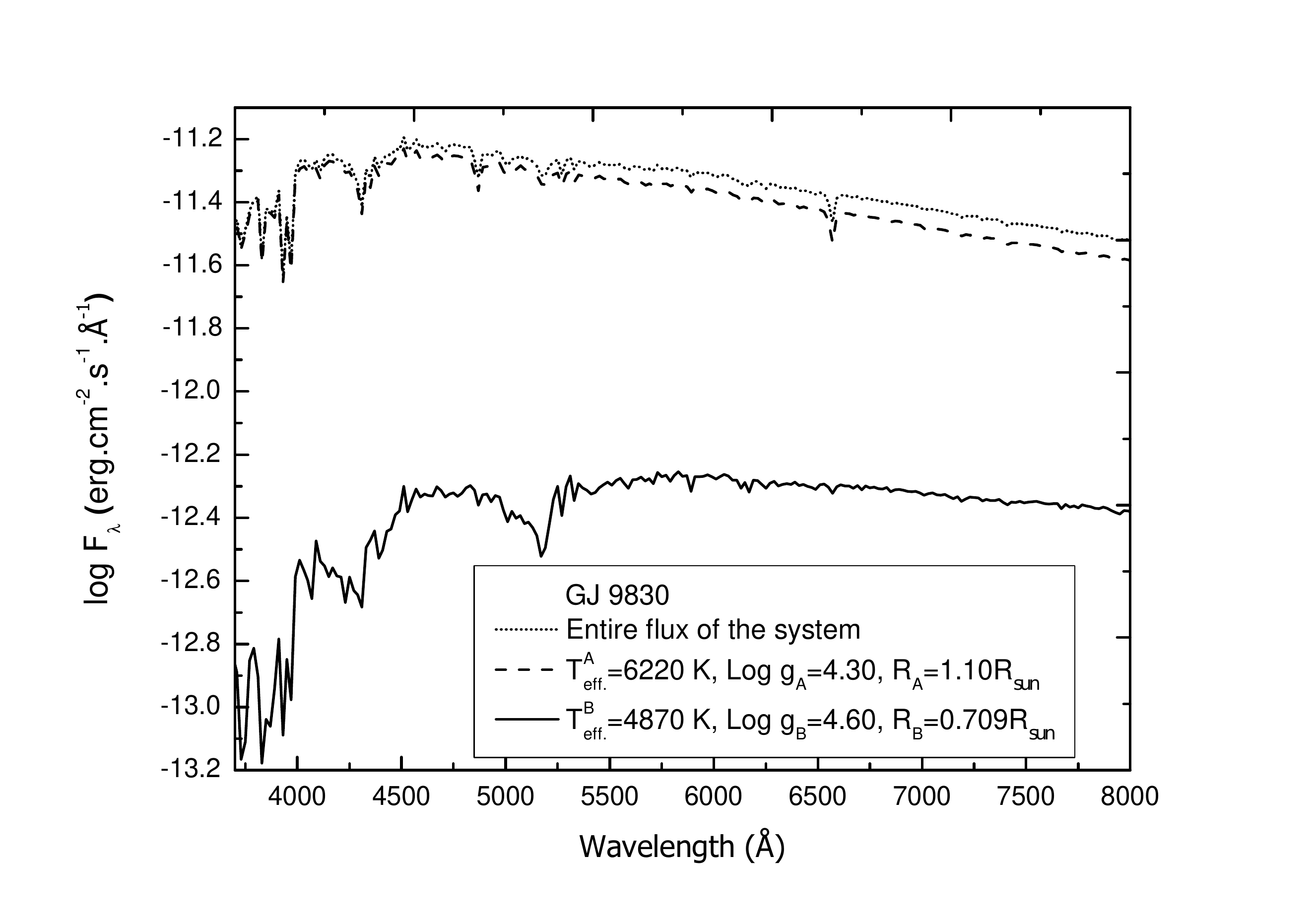}
	\caption{The entire and individual synthetic SEDs of the system by using Al-Wardat's method depending on Kurucz line blanketed models, as it would be if measured from outside the earth's atmosphere at a distance of 34.27 pc from the star.\label{fig1}}
\end{figure}

The best agreement between observed and synthetic photometry is achieved at a set of the stellar parameters of the individual components of the system ($T_{\rm eff.}$, $\log\rm g$, $R$ and $d$ ) which are showed in Fig.~\ref{fig1} and  listed in Table~\ref{tablef1}.

The errors of the individual radii components have double-checked by using the following equation:
\begin{eqnarray}
	\label{eq3412}
	\sigma_{R} \approx\pm R\sqrt{(\frac{\sigma_{M_{\rm bol}}}{5 \log e})^2+4(\frac{\sigma_{T_{\rm eff}}}{ T_{\rm eff} })^2}
\end{eqnarray}
Here $M_{\rm bol}$ is the bolometric magnitudes for each system's component.

Based on the ultimate radii and effective temperatures of the system, the stellar luminosities and bolometric magnitudes along with their errors are listed in Table~\ref{tablef1} and the spectral type of GJ9830 A is
found to be F7.5V and GJ9830 B to be K3.5V.

To place the individual components of the system on the theoretical Hertzsprung-Russell (H-R)
diagram and estimate the age and total mass of the system, we use the values of log $\rm L/\rm L_\odot$ and log $T_{\rm eff}$ based on the evolutionary tracks of \citep{2000yCat..41410371G} (see Fig.\ref{a25}) and
isochrones of \citep{2000A&AS..141..371G} (see Fig.\ref{a26}). The positions of the components of the stars in these diagrams lead to theoretical estimates of their masses and ages. As a result, the individual masses of the system are $\mathcal{M}^{A}=1.18\pm0.10 \mathcal{M}_\odot$ and $\mathcal{M}^{B}=0.75\pm0.08 \mathcal{M}_\odot$ for the primary and secondary components, respectively with a system age of $1.40\pm0.50$ Gyr.
\section {Results and discussion}\label{4}

Table~\ref{tabb} shows the results of the accurate orbital parameters of the close binary system, GJ 9830, which are shown in Figure~\ref{figor}. The rms of the binary system are $0.\degr85$ and $0.''006$. \cite{2007AstBu..62..339B} estimated the orbital parameters by using \cite{1977ApJ...214L.133M} method, which are acutely agreement with our study in a certain of parameters despite of availability of more relative positions measurements in our case. At the same time, there is the best agreement between results of Tokovinin's dynamical and Al-Wardat's method in terms of total mass of the binary system. Moreover, the residuals, $\Delta \theta$ and $\Delta \rho$ of the binary system are shown in Table~\ref{tab}.

\begin{table*}[h]
	\begin{center}
		\caption{Orbits, total mass, and quality controls published for the GJ\,9830 system, compared with the orbital solution calculated in this work.}
		\label{tabb}
		\begin{tabular}{cccc}
			\noalign{\smallskip}
			\hline\hline
			\noalign{\smallskip}
			&	&\multicolumn{2}{c}{System GJ 9830} \\
			\cline{3-4}
			\noalign{\smallskip}
			Parameters	& Units & \cite{2007AstBu..62..339B} & This work\\
			\hline
				\noalign{\smallskip}
			$\rm P$  & [yr]       & $15.70\pm 0.23$ &   $16.368\pm 0.032$
			\\
			$\rm T_0$  & [yr]   & $2005.49 \pm 0.01$ &   $2005.662 \pm 0.021$
			\\
			$\rm e$   & -  & $0.536 \pm 0.007$ &    $0.537\pm 0.006$
			\\
			$\rm a $ & [arcsec] & $0.220\pm 0.002 $ &    $0.225\pm 0.002 $
			\\
			$\rm i $  & [deg]    & $ 75.1\pm 0.4$  &     $ 74.94\pm 0.220$
			\\
			$\rm \Omega$   & [deg] & $141.5 \pm 0.3$  &   $141.50\pm 0.18$
			\\
			$\rm \omega$  & [deg] & $89.5\pm 0.8$   &   $89.50\pm 0.180$
			\\
			$\mathcal{M}_T $ & [$\mathcal{M}_\odot$] & $1.56\pm0.18$  &     $1.75\pm0.06$
			\\
			rms ($\theta$)& [deg] & 3.25 &  0.85\\
			rms ($\rho$)& [arcsec] & 0.003 &  0.006\\
			$\pi_{Hip}$  & [mas]   & $30.24\pm1.12$ $^{a}$ &  $29.178\pm0.186$ $^{b}$
			\\
			\hline\hline
			\noalign{\smallskip}
		\end{tabular}
		\\
		\medskip
		$^{a}$ The old parallax \cite{1997yCat.1239....0E}, $^{b}$ The Gaia parallax  \cite{2018yCat.1345....0G}.
	\end{center}
\end{table*}

Table~\ref{tab42} shows the results of the calculated magnitudes and color  indices of the entire
synthetic system and individual components of the system GJ 9830. Table~\ref{synth3} shows the best agreement between the entire synthetic magnitudes and color indices  with the observed ones within three photometric systems: Johnson-Cousins, Str\"{o}mgren and Tycho. This led to the most important indication for the reliability of the calculated physical parameters of the close binary systems, GJ9830, listed in Table~\ref{tablef1}.
\begin{table}
	\small
	\begin{center}
		\caption{ Magnitudes and color indices of the composed synthetic spectrum and individual components of GJ\,9830.}
		\label{tab42}
		\begin{tabular}{lcccc}
			\noalign{\smallskip}
			\hline\hline
			\noalign{\smallskip}
			Sys. & Filter & Entire Synth.& GJ\,9830 & GJ\,9830\\
			&     & $\sigma=\pm0.03$&   A    &     B      \\
			\hline
			\noalign{\smallskip}
			Joh-          & $U$ & 7.84 & 7.88 & 11.51 \\
			Cou.          & $B$ & 7.73  &  7.81 &  10.72  \\
			& $V$ & 7.14 &  7.25 &  9.72 \\
			& $R$ & 6.81  &  6.95 & 9.14  \\
			&$U-B$& 0.10  & 0.07 & 0.78 \\
			&$B-V$& 0.59  &  0.56 &  1.00 \\
			&$V-R$& 0.33  &  0.31 & 0.58 \\
			\hline
			\noalign{\smallskip}
			Str\"{o}m.    & $u$ & 9.01 & 9.05 &  12.73  \\
			& $v$ & 8.06 & 8.12  & 11.31  \\
			& $b$ & 7.48 & 7.57 &  10.24 \\
			&  $y$& 7.11 & 7.22 & 9.66  \\
			&$u-v$& 0.95 & 0.93 & 1.42 \\
			&$v-b$& 0.58 & 0.55 & 1.08 \\
			&$b-y$& 0.36 & 0.34 & 0.58 \\
			\hline
			\noalign{\smallskip}
			Tycho       &$B_T$  & 7.87   & 7.94& 11.0   \\
			&$V_T$  & 7.21   & 7.32 & 9.83 \\
			&$B_T-V_T$& 0.66 & 0.62 & 1.17\\
			\hline\hline
			\noalign{\smallskip}
		\end{tabular}
	\end{center}
\end{table}
\begin{table}
	\small
	\begin{center}
		\caption{Comparison between the entire synthetic and entire observational
			magnitudes, colours and magnitude differences for the system.} \label{synth3}
		\begin{tabular}{ccc}
			\noalign{\smallskip}
			\hline\hline
			\noalign{\smallskip}
			&\multicolumn{2}{c}{GJ9830}\\
			\cline{2-3}
			\noalign{\smallskip}
			Filter	& Entire obs. $^a$ & Entire synth.$^b$(This work) \\
			\hline
			\noalign{\smallskip}
			$V_{J}$ & $7\fm14$            & $7\fm14\pm0.03$\\
			$B_J$& $7\fm72$               & $7\fm73\pm0.03$  \\
			$(B-V)_{J}$&$ 0\fm59\pm0.008$ &$ 0\fm59\pm0.04$\\
			$(b-y)_{S}$&$ 0\fm40\pm0.002$ &$ 0\fm36\pm0.04$\\
			$(v-b)_{S}$&$ 0\fm58\pm0.002$ &$ 0\fm58\pm0.04$\\
			$(u-v)_{S}$&$ 0\fm86\pm0.007$ &$ 0\fm95\pm0.04$\\
			$B_T$  & $7\fm86\pm0.007$      &$7\fm87\pm0.03$\\
			$V_T$  & $7\fm23\pm0.006$      &$7\fm21\pm0.03$\\
			\hline
			$\triangle m$  &$ 2\fm47^{c}\pm0.07$  &$ 2\fm47^{d}$\\
			\hline\hline \noalign{\smallskip}
		\end{tabular}\\
		\textbf{Notes.}
		$^a$ The real observations (Table~\ref{tabl0}).\\
		$^{b}$ Entire synthetic work of GJ 9830  (Table~\ref{tab42}).\\
		$^{c}$ Average magnitude differences for all $\triangle m$ under the 541-562 nm V-band filter  (Table~\ref{taba1}).\\
		$^{d}$  $\triangle m$ = $V^{B}_{J}$-$V^{A}_{J}$ for the system (Table~\ref{tab42}). \\

	\end{center}
\end{table}
\begin{table}
	\small
	\begin{center}
		\caption{The ultimate  stellar parameters of the components of the system GJ9830.} \label{tablef1}
		\begin{tabular}{cccc}
			\noalign{\smallskip}
			\hline\hline
			\noalign{\smallskip}
			& &\multicolumn{2}{c}{GJ9830}\\
			\cline{3-4}
			\noalign{\smallskip}
			Parameter & Units	& GJ9830 A & GJ9830 B  \\
			\hline
			\noalign{\smallskip}
			$\rm T_{\rm eff.}$  &[K] & $6220\pm100$ & $4870\pm100$ \\
			R  & [R$_{\odot}$] & $1.10\pm0.08$ & $0.709\pm0.07$ \\
			$\log\rm g$& [cgs]&$4.30\pm0.12$ & $4.60\pm0.11$ \\
			$\rm L $  & [$\rm L_\odot$] & $1.63\pm0.05 $  & $0.25\pm0.04$ \\
			$\rm M_{bol}$  & [mag] &  $4.22\pm0.21$ & $6.26\pm0.20$ \\
			$ \mathcal{M}_\odot$$^{1}$   &[$\mathcal{M}_\odot$]&  $1.18 \pm0.10$ & $0.75 \pm0.08$ \\
			Sp. Type$^{2}$ &  & F7.5V  &  K3.5V \\
			\hline\noalign{\smallskip}
			\multicolumn{1}{c}{Parallax $^{3}$ }  & [mas]&  \multicolumn{2}{c}{$29.178 \pm 0.186 $} \\
			\multicolumn{1}{c}{Age $^{4}$ }& [Gyr] & \multicolumn{2}{c}{ $1.40\pm 0.50$} \\
			\hline\hline\noalign{\smallskip}
		\end{tabular}\\
		\textbf{ Notes.} $^{1}${Depending on the evolutionary tracks of~\cite{2000yCat..41410371G} (Fig.~\ref{a25})},\\
		$^{2}${Using the tables of~\citep{1992adps.book.....L,2005oasp.book.....G}.\\
			$^{3}$ \cite{2018yCat.1345....0G}.\\
			$^{4}${Depending on the the isochrones of
				different metallicities of~\cite{2000A&AS..141..371G} ( Figs.~\ref{a27}}).}
	\end{center}
\end{table}

The results of the apparent magnitudes $m_v$ from synthetic photometry are found to be completely similar to those from the observed photometry of the binary system. At the same time, the difference between synthetic and observed valuse of magnitudes and colours indices in the different photometrical systems of the binary system is less than 0.04 $\sigma$. The agreement between these values indicates an accuracy of the method and an indication for the reliability of the calculated stellar parameters of the system.

Fig.~\ref{fig1} show the entire and individual synthetic spectral energy distributions of the close binary system, GJ9830, based on the calculated stellar parameters and on the revised distance of the system from \textit{Gaia}
astrometric mission \citep{2018yCat.1345....0G}.

\begin{figure*}
	\centering
	\includegraphics[width=0.8\columnwidth]{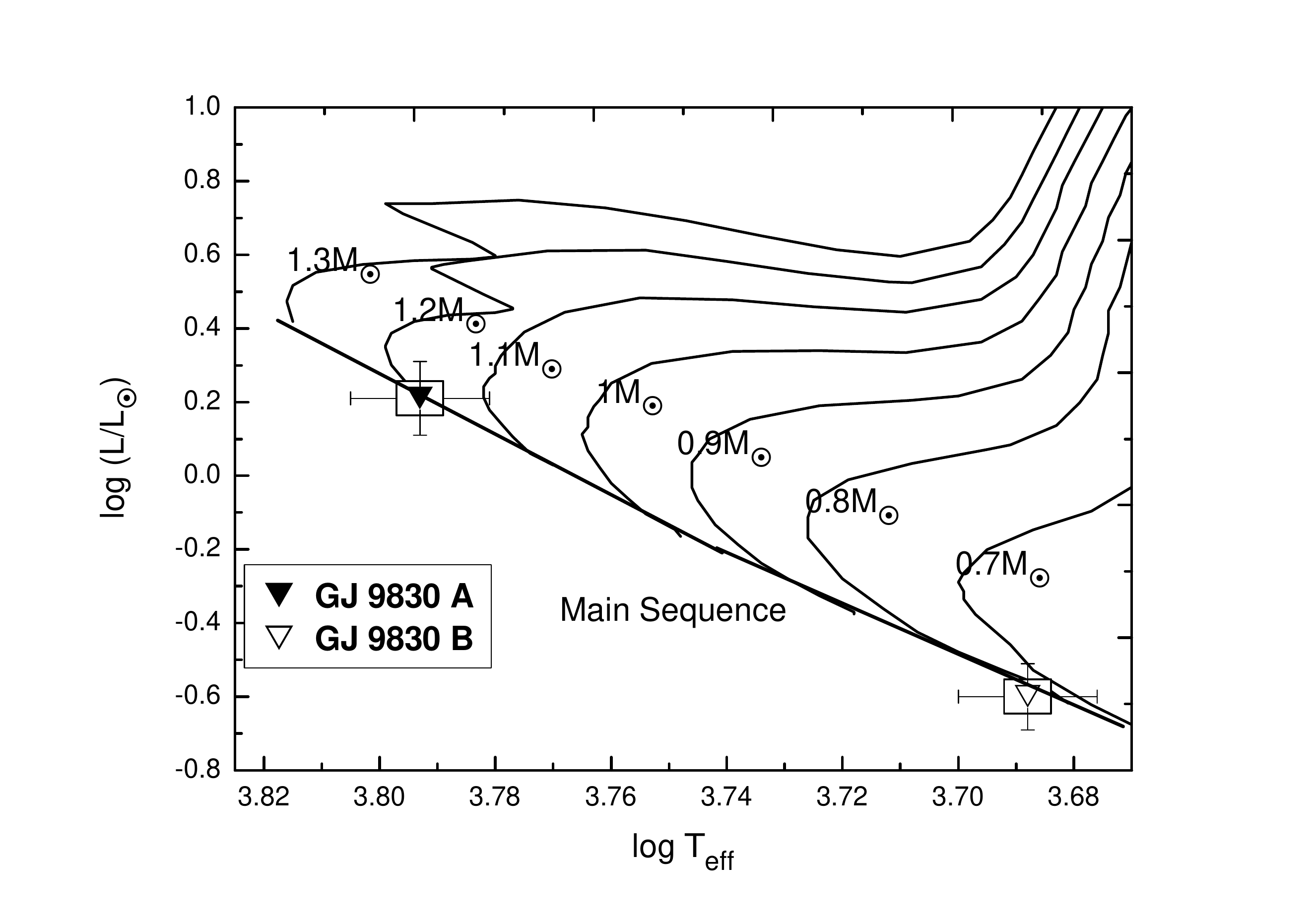}
	\caption{The evolutionary tracks of both components of GJ\,9830 on the H-R diagram of masses ( 0.7, 0.8,...., 1.3 $\rm\,M_\odot$). The evolutionary tracks were taken from~ \cite{2000yCat..41410371G}.} \label{a25}
\end{figure*}

To estimate the stellar masses and ages of the system, we used \cite{2000yCat..41410371G}'s theoretical H-R diagram with evolution tracks, and the isochrones given by \cite{2000A&AS..141..371G}, respectively. The GJ 9830's components on the evolutionary tracks of \cite{2000yCat..41410371G} (Fig.~\ref{a25}) belong to the main-sequence stars. According to that, the stellar masses of the binary system were computed by using two different methods- Al-wardat's and Tokovinin's method, based on the Gaia parallax \cite{2018yCat.1345....0G}. The former gave $ \mathcal{M}^{A}=1.18\pm0.10\, \mathcal{M}_\odot$, $ \mathcal{M}^{B}=0.75\pm0.08\, \mathcal{M}_\odot$ for the primary and secondary components, respectively, while the latter gave $1.75\pm0.06\rm\, \mathcal{M}_\odot$. The total mass by using Al-wardat's complex method are found to be similar to those from Tokovinin's method. This revealed the accuracy of the used methods for the binary system.
 .
\begin{figure*}
	\centering
	\includegraphics[width=0.8\columnwidth]{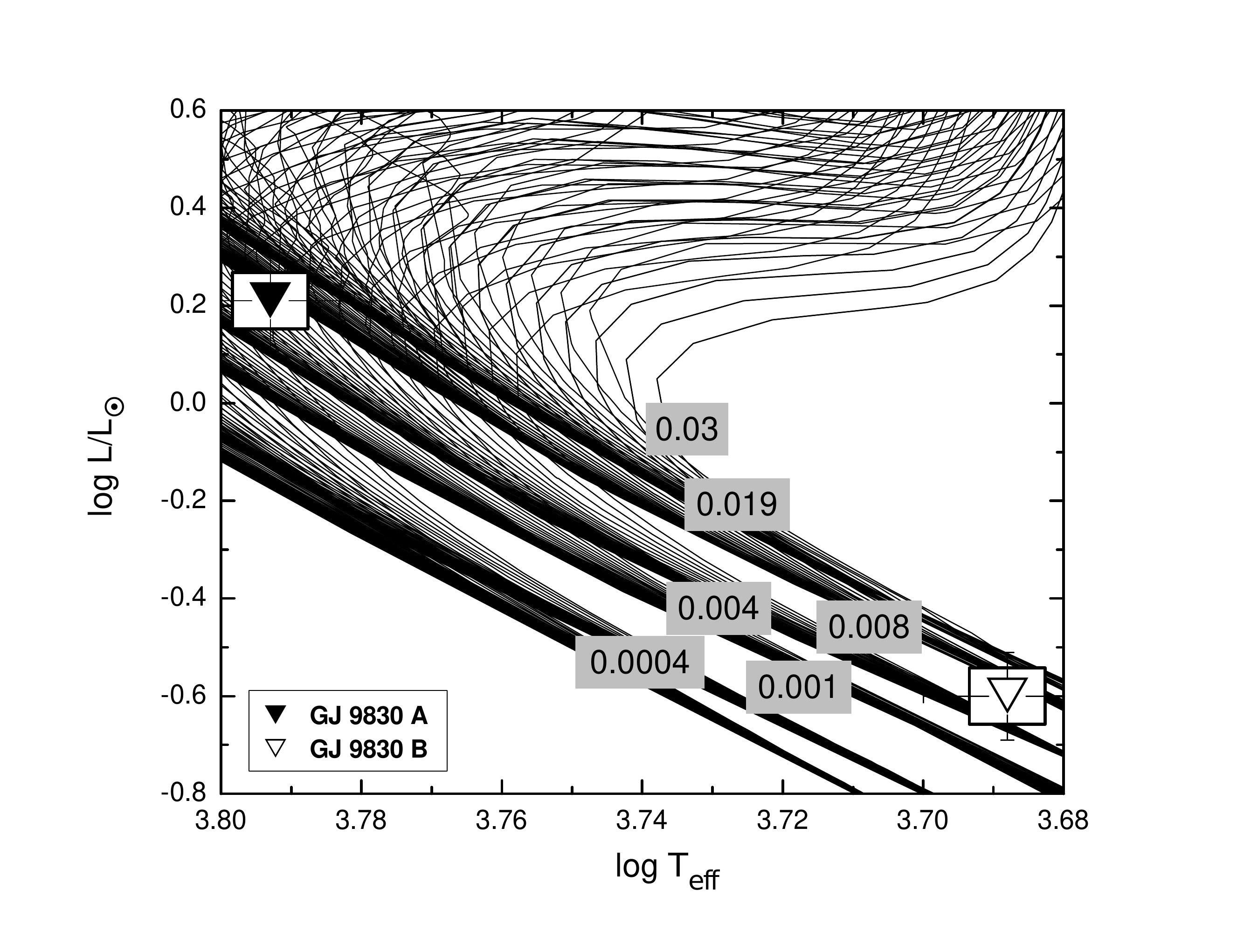}
	\caption{The isochrones for both components of GJ\,9830 on the H-R diagram for low- and intermediate-mass: from $0.15 $ to $7.0 \rm\,M_\odot$ stars of
		different metallicities. The isochrones were taken from~\cite{2000A&AS..141..371G}.} \label{a26}
\end{figure*}

Fig.~\ref{a26} shows the initial chemical compositions [Z = 0.0004, Y = 0.23], [Z = 0.001, Y = 0.23], [Z = 0.004, Y =0.24], [Z = 0.008, Y = 0.25], [Z = 0.019, Y = 0.273] (solar composition), and [Z = 0.03, Y = 0.30]. Acording to the positions of the components of the binary system on tracks, the helium and metal mass fractions, Y and Z, respectively are [Z = 0.019, Y = 0.273] (solar composition).

\begin{figure*}
	\centering
	\includegraphics[width=0.83\columnwidth]{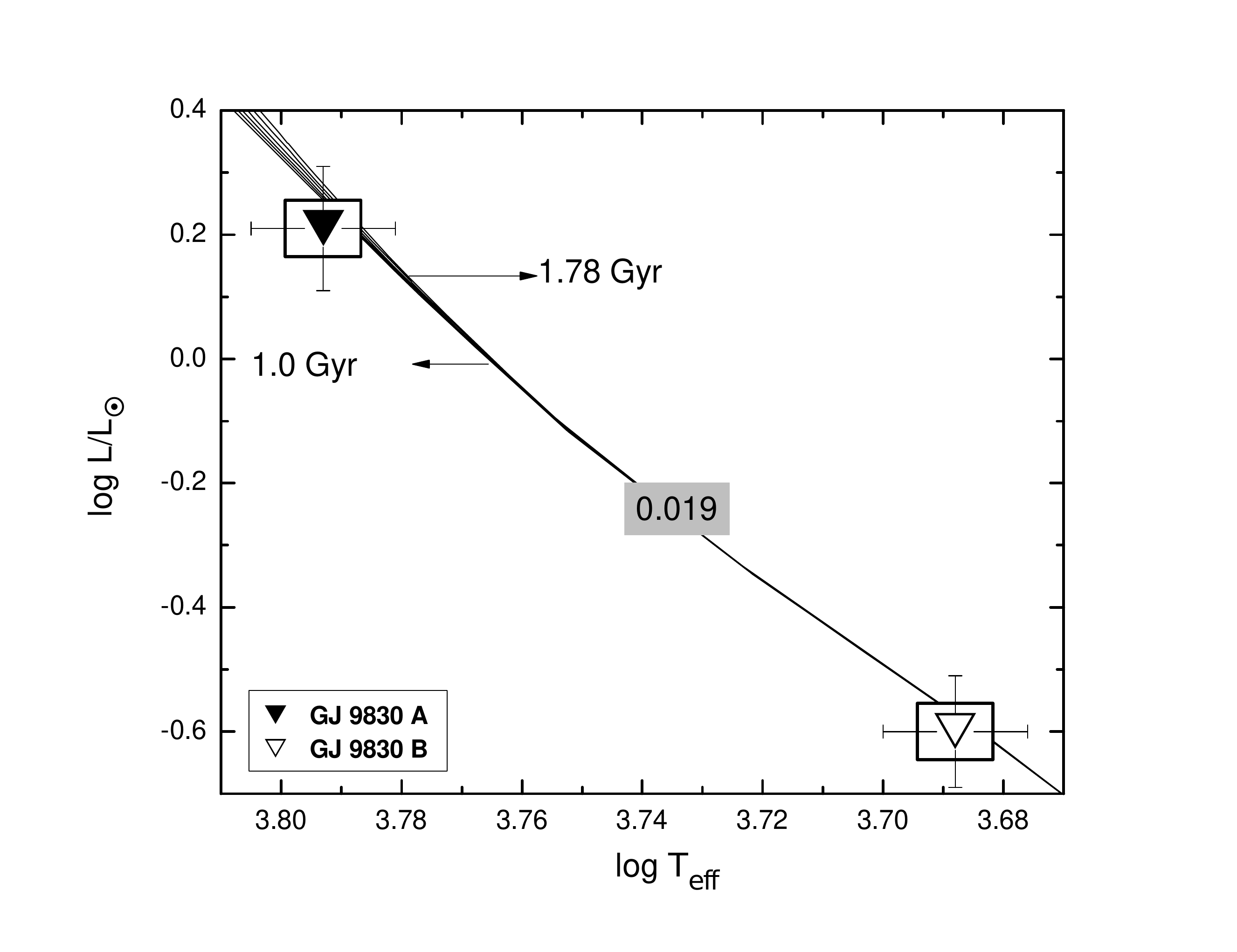}
	\caption{The isochrones for both components of HIP\,9830 on the H-R diagram for low- and intermediate-mass: from $0.15 $ to $0.99 \rm\,M_\odot$ , and for the compositions [Z=0.019, Y=0.273] stars of different metallicities. The isochrones were taken from~\cite{2000A&AS..141..371G}.} \label{a27}
\end{figure*}

Fig~\ref{a27} shows the components of the GJ\,9830 system on isochrones.  
 It is clear from the parameters of the system's components and their positions on the evolutionary tracks that both components belong to the young solar type main sequence stars, with age around $1.40\pm0.50$\,Gyr.   
 
 Depending on the formation theories,   fragmentation
is suggested as the most likely process for the formation of the system. 
Where \cite{1994MNRAS.269..837B} concludes that fragmentation of a rotating disk
around an incipient central protostar is possible, as long as
there is continuing infall.  \cite{2001IAUS..200.....Z} pointed out that
hierarchical  fragmentation during rotational collapse has been
invoked to produce binaries and multiple systems.

\section{Conclusions}
By using Al-Wardat's method for analysing close visual binary systems which employs  Kurucz {\fontfamily{cmtt}\selectfont ATLAS9} line-blanketed plane-parallel model atmospheres in constructing the synthetic SED and applying the synthetic photometry on the synthetic SED, we were able to evaluate the physical parameters of the Main-Sequence system, GJ\,9830. The present analysis shows that the binary GJ\,9830 belong to a class of the Main-Sequence systems. The results from synthetic photometry are found to be similar to those from the observed ones, which revealed the accuracy of the used method and led to estimate the best stellar parameters for the binary system.

The orbital parameters of the system calculated properly by using Tokovinin's dynamical method. These parameters gave accurate total mass of the binary system as  $1.75\pm0.06\rm\, \mathcal{M}_\odot$  based on the new parallax from Gaia and on revised orbits of the binary system.

The positions of the  components of the system have been shown with a broad way on the evolutionary tracks and isochrones. The spectral types of the components of GJ\,9830 are catalogued
 as F7.5V and K3.5V for the primary and secondary components of the system, respectively with an age of $1.40\pm 0.50$\,Gyr.
The evolutionary tracks and isochrones of the system's components are discussed, and the fragmentation process is suggested as the most likely process for the formation of the system.

\begin{acknowledgements}
	This research has made use of SAO/NASA, SIMBAD database, Fourth Catalog of Interferometric Measurements of Binary Stars, IPAC data systems, ORBIT code and CHORIZOS code of photometric and spectrophotometric data analysis.
\end{acknowledgements}

\bibliographystyle{raa}
\bibliography{references}

\end{document}